\documentclass[10pt,pra,twocolumn,floatfix]{revtex4}
\usepackage{amsfonts}
\usepackage{amssymb}
\usepackage{amsmath}
\usepackage[dvips]{graphicx}

\begin{document}

\title{Generating quantum correlated twin beams by four-wave mixing in hot cesium vapor}
\author{Rong Ma$^{1}$}
\author{Wei Liu$^{1}$}
\author{Zhongzhong Qin$^{1,2}$}
\email{zzqin@sxu.edu.cn}
\author{Xiaojun Jia$^{1,2}$}
\author{Jiangrui Gao$^{1,2}$}
\affiliation{$^1$State Key Laboratory of Quantum Optics and Quantum Optics Devices,
Institute of Opto-Electronics, Shanxi University, Taiyuan 030006, People's
Republic of China\\
$^2$Collaborative Innovation Center of Extreme Optics, Shanxi University,
Taiyuan, Shanxi 030006, People's Republic of China\\
}

\begin{abstract}
Using a nondegenerate four-wave mixing process based on a double-$\Lambda$ scheme in hot cesium vapor, we generate quantum correlated twin beams with a maximum
intensity-difference squeezing of 6.5 dB. The substantially improved squeezing can be mainly attributed to very good frequency and phase-difference stability between the pump and probe beams in our experiment.
Intensity-difference squeezing can be observed within a wide experimental parameter range, which guarantees its robust generation. Since this scheme produces multi-spatial-mode twin beams at the
Cs $D_{1}$ line, it is of interest for experiments involving quantum imaging and coherent interfaces between atomic and solid-state systems.
\end{abstract}

\maketitle

\section{Introduction}
Quantum correlation and entanglement are significant for both fundamental tests of quantum physics \cite{EPR,Bell} and applications in
future quantum technologies, such as quantum metrology \cite{GravitationalWave}, quantum imaging \cite{QuantumImaging}, and quantum information processing \cite{BraunsteinRMP}.
The standard technique to generate quantum correlated beams and continuous-variable entangled states is by parametric down-conversion in a nonlinear crystal,
with an optical parametric oscillator (OPO) or optical parametric amplifier. While very large amounts of quantum noise reduction have been achieved in this way \cite{LauratOL,JiaOE},
the central frequency and linewidth of the generated nonclassical states usually do not naturally match the transitions of matters, such as atoms and solid-state systems,
which limits their applications in  light-matter interactions.

On the other hand, the first experimental demonstration of a squeezed state of light was realized using a four-wave mixing (FWM)
process in sodium vapor \cite{SlusherPRL}. Since then, many groups have demonstrated squeezing in atomic vapor under a variety of different configurations
\cite{AtomsSqueezing1,AtomsSqueezing2,AtomsSqueezing3,AtomsSqueezing4,AtomsSqueezing5,AtomsSqueezing6,AtomsSqueezing7}.
However, the decoherence effect caused by spontaneous emission and absorption in atomic vapor limits the level of quantum noise reduction to no more than 2.2 dB \cite{AtomsSqueezing7}.
The nondegenerate FWM process in a double-$\Lambda$ scheme was recognized as a possible work-around for these limitations,
in which ground-state coherence based on coherent population trapping and electromagnetically induced transparency can reduce or eliminate
spontaneous emission noise \cite{DoubleLambda1,DoubleLambda2}. Recently, it was shown by Lett's group \cite{PooserOE,LettPRA} and several other groups  \cite{QinOL,GlorieuxArxiv,JasperseOE} that the FWM process based on a double-$\Lambda$ scheme
in hot Rb vapor is an efficient way to generate quantum correlated twin beams. The highest degrees of intensity-difference squeezing obtained by several different groups are in the range of 7.0 to 9.2 dB \cite{PooserOE,QinOL,GlorieuxArxiv},
which approaches the best value reported for OPO \cite{LauratOL}. This system has proven to be very successful for a variety of applications \cite{EntangledImages,TunableDelay,SU11NC,QinPRL,QinAPL,MacRaPRL,QinLight,TravisOL}.

\begin{figure}[b]
\includegraphics[width=8cm]{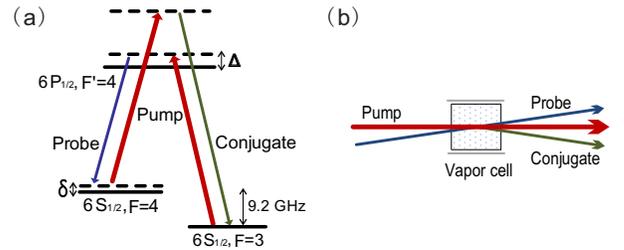}
\caption{Experimental layout for generating quantum correlated twin beams. (a) Double-$\Lambda$ scheme in the $D_{1}$ line of $^{133}$Cs. $\Delta$ is one-photon detuning and $\delta$ is two-photon detuning. (b) Schematic of the FWM process.}
\end{figure}

\begin{figure*}[htbp]
\includegraphics[width=17cm]{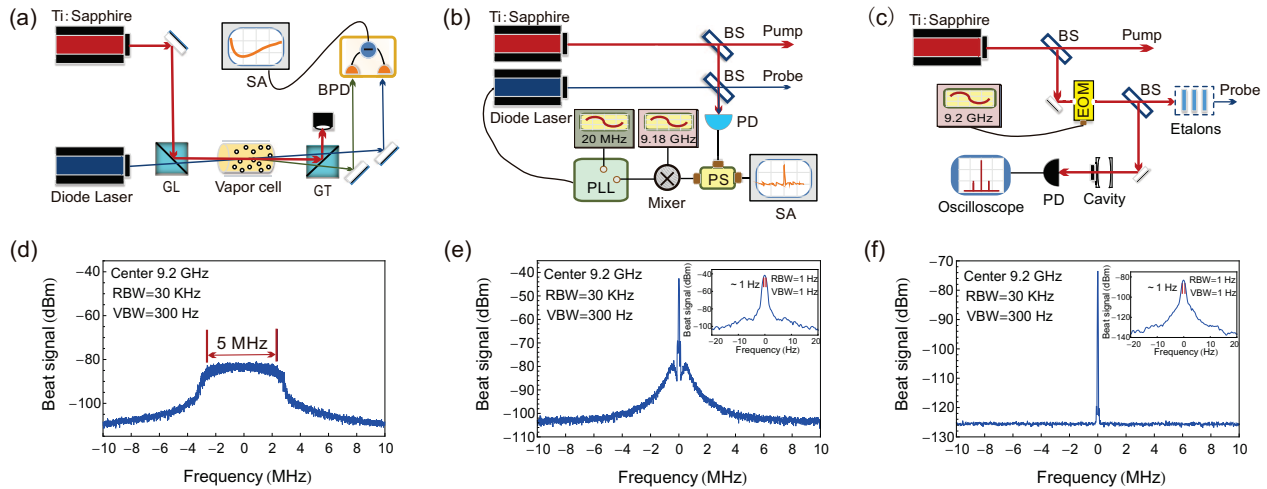}
\caption{Experimental setup for generating and detecting quantum correlated twin beams with three different methods to generate the probe beam. (a) A Ti:sapphire laser and a diode laser are used as pump and probe beams, respectively. GL: Glan-laser polarizer;
GT: Glan-Thompson polarizer; BPD: balanced photodetector; SA: spectrum analyzer. (b) The same lasers are used, and the PLL works in the meanwhile. BS: beam splitter; PD: photodetector; PS: power splitter; PLL: phase-locked loop.
(c) An electro-optic modulator (EOM) is used to generate the probe beam. (d), (e), and (f) Corresponding beat spectrum of the pump and probe beams for the three methods shown in (a), (b), and (c), respectively. Insets of (e) and (f) show the beat spectrum in a 40-Hz span.}
\end{figure*}

Twin beams have also been experimentally generated based on the FWM process in other hot alkali vapor, e.g., cesium \cite{CsTwinBeams} and potassium \cite{KTwinBeams}, as well as
in metastable helium at room temperature \cite{HeTwinBeams}. Cesium offers certain advantages, e.g., the quantum correlation at the $^{133}$Cs $D_{1}$ line lies well within the
wavelength regime of the exciton emission from InAs quantum dots \cite{QuantumDots}, which provides a potential resource for coherent interfaces between atomic and solid-state systems.
The highest degree of intensity-difference squeezing of twin beams based on the FWM process in hot cesium vapor is 2.5 dB \cite{CsTwinBeams}. In Ref. \cite{CsTwinBeams}, the authors pointed out the experimental difficulty in getting high quantum correlation in cesium vapor
because of larger hyperfine splitting of the ground states compared with Rb, as well as experimental limitations such as laser frequency stability, the vapor cell's transmission efficiency, and mechanical vibration of the system.
In this paper, we substantially improve the squeezing to 6.5 dB, which can be mainly attributed to very good frequency and phase-difference stability between the pump and probe beams in our experiment.
Our work shows it is possible to produce narrowband quantum correlated twin beams with similar double-$\Lambda$ schemes at a particular wavelength \cite{KTwinBeams,HeTwinBeams}.

\section{Experimental setup}
As shown in Fig. 1(a), the $^{133}$Cs $D_{1}$ line is used to form the double-$\Lambda$ level structure with an excited level
($6P_{1/2}, F'=4$) and two ground levels ($6S_{1/2}, F=3$ and $F=4$). A weak probe beam with an intensity of $I_{0}$ crosses
with a strong pump beam inside a hot Cs vapor cell at a small angle [Fig. 1(b)]. After the FWM process, the intensity of the probe beam
is amplified to $GI_{0}$. In the meanwhile, another conjugate beam is generated on the other side of the pump beam with an intensity of $(G-1)I_{0}$.
The FWM process relies on the double-$\Lambda$ scheme in which two pump photons are simultaneously converted to one probe photon and one conjugate photon.
As a result, the relative intensity difference of the probe and conjugate beams is squeezed
compared with the corresponding shot-noise limit (SNL) by an amount of $1/(2G-1)$.

To produce quantum correlated twin beams efficiently, the pump-probe frequency difference is required to be approximately the ground-state hyperfine splitting (9.2 GHz for $^{133}$Cs).
There are several methods to achieve two laser beams with several gigahertz of frequency separation: (i) using two independent free-running lasers, (ii) using a phase-locked loop (PLL) to lock the frequency
and phase difference of two lasers \cite{LvovskyPLL}, (iii) using the diffraction effect of an acousto-optic modulator (AOM) to generate a laser beam with a certain frequency difference \cite{LettPRA,QinOL,GlorieuxArxiv,JasperseOE}, and (iv) generating a sideband with a certain frequency difference by the use of an electro-optic modulator (EOM) \cite{EOMSideband}.

We generate quantum correlated twin beams based on a FWM process in cesium vapor with
the 9.2-GHz frequency-difference probe beam achieved using methods (i), (ii), and (iv). Method (iii) is not chosen because an AOM which has
a 9.2-GHz frequency-shift ability is not commercially available. Furthermore, the diffraction efficiency of an AOM with a 3-GHz frequency shift is on the order
of 1\% \cite{LettPRA,QinOL,GlorieuxArxiv,JasperseOE}, so achieving a 9.2-GHz frequency shift with multiple cascaded AOMs is impractical.

\begin{figure}[t]
\includegraphics[width=8.5cm]{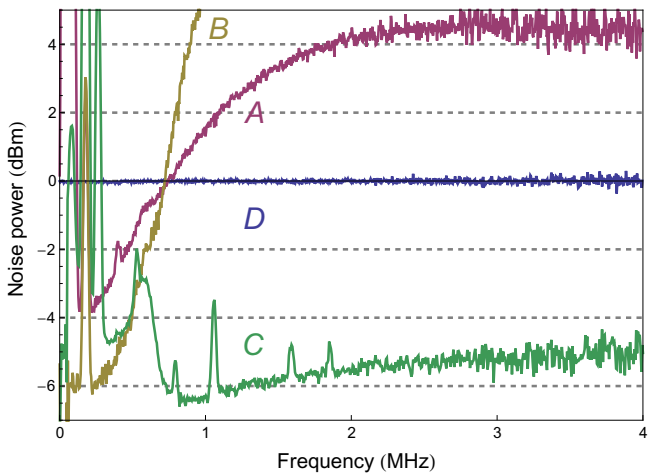}
\caption{Intensity-difference noise spectra of the quantum correlated twin beams with the probe beam generated using another laser (trace $A$), PLL (trace $B$), and the EOM sideband (trace $C$). Trace $D$ gives corresponding SNL.
The electronic noise floor and background noise are both 10 dB below the corresponding SNL at 1 MHz and have been subtracted from all of the traces.The sharp peaks on traces $A$, $B$, and $C$ can be attributed to classical noise from our lasers.}
\end{figure}

\subsection{Using two independent lasers}
First, we use a Ti:sapphire laser and a diode laser as the pump beam and probe beam, respectively [Fig. 2(a)]. The Ti:sapphire laser is tuned about 1.6 GHz
to the blue of $^{133}$Cs ($6S_{1/2}, F=3\rightarrow6P_{1/2}, F'=4$) with a total power of 1 W. Beams from the Ti:sapphire laser and the diode laser are mixed at a beam splitter,
and a total optical power of 5 mW is focused onto a high-bandwidth photodetector [shown in Fig. 2(b)]. As shown in Fig. 2(d), the beat signal of these two lasers, with a full width at half maximum (FWHM) of around 5 MHz,
is monitored on a spectrum analyzer (SA) to ensure its central frequency is equal to the ground-state hyperfine splitting (two-photon detuning $\delta=0$ MHz).

As shown in Fig. 2(a), by choosing vertical polarization for the pump and horizontal polarization for the probe, they can be combined in a Glan-laser (GL) polarizer.
The beams then cross each other at an angle of 6 mrad in the center of the cesium vapor cell. The vapor cell is 25 mm long and temperature stabilized at $112 ^{\circ}$C.
The windows of the vapor cell are antireflection coated at 895 nm on both faces, resulting in a transmissivity for the far-detuned probe beam of 98\% per window.
The pump beam and the probe beam are focused with waists of 560 and 300 $\mu$m (1/$e^{2}$ radius), respectively, at the crossing point to ensure that they overlap over almost the full length of the cell.
 After the vapor cell, a Glan-Thompson (GT) polarizer with an extinction ratio of $10^{5}$:1 is used to filter out the pump beam. The amplified probe and the generated conjugate beams are directly sent into the
 two ports of a balanced photodetector (BPD) with a gain of $10^{5}$ V/A and a quantum efficiency of 98\%. The output of the BPD is sent to a SA
 with a resolution bandwidth (RBW) of 30 kHz and a video bandwidth (VBW) of 300 Hz. To measure the SNL, a coherent laser beam, whose power is equivalent to the total power of the probe and conjugate beams,
 is split into two beams using a 50:50 beam splitter and sent to the BPD.

\begin{figure}[t]
\includegraphics[width=8.5cm]{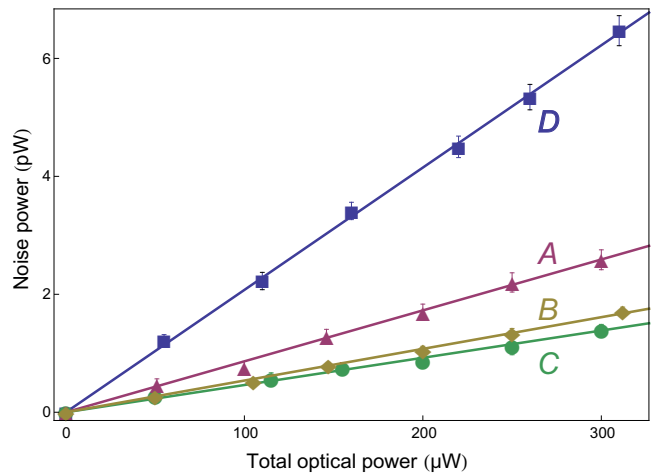}
\caption{Intensity-difference noise power versus total optical power for twin beams using two independent lasers (curve $A$), PLL (curve $B$), and EOM (curve $C$), and for a coherent beam (curve $D$). The electronic noise and background noise have been subtracted from all the data points.
Error bars for experimental data represent $\pm1$ standard deviation and are obtained based on the statistics of the measured data.}
\end{figure}

 \subsection{Using a PLL}
 The basic schematic for our PLL is shown in Fig. 2(b). A power splitter splits the beat signal between a SA and the PLL. Two reference frequencies, one 9.18 GHz and the other 20 MHz,
 are used to  produce the error signal for the diode laser, closing the feedback loop. A commercial fast analog linewidth controller allows fast current injection modulation of the diode laser up to tens of megahertz.

Shown in Fig. 2(e) is the beat spectrum with a span of 20 MHz when the PLL works. The inset shows the beat spectrum in a 40-Hz span. It clearly shows the FWHM of the beat signal is around 1 Hz, which indicates very good frequency and phase-difference stability between the
 pump and probe beams.

 \subsection{Using an EOM}
As shown in Fig. 2(c), we use an EOM to produce optical sidebands at $\pm9.2$ GHz from the carrier frequency of the Ti:sapphire laser. The EOM is driven by a signal generator and a power amplifier with a radio-frequency
power of 34 dBm. The tuning range of the EOM is $\pm50$ MHz centered at 9.2 GHz. The relative optical power in the first-order sidebands and carrier is monitored by a scanning cavity and amounts to 10\%, 10\%, and 80\%, respectively.
To select the probe frequency component (-1$^{st}$ order sideband) from the carrier and two sidebands, three successive temperature-stabilized etalons with a finesse of around 60 are used, which give a total transmissivity
of around 80\%. The 9.2-GHz frequency-shift efficiency of 8\% is high enough that no further amplification of the probe beam is needed.

Shown in Fig. 2(f) is the beat spectrum of the -1$^{st}$ order sideband and the carrier frequency with a span of 20 MHz. The inset of Fig. 2(f) shows the beat spectrum in a 40-Hz span. The FWHM of the beat signal is also around 1 Hz.

\begin{figure*}[tbph]
\includegraphics[width=15cm]{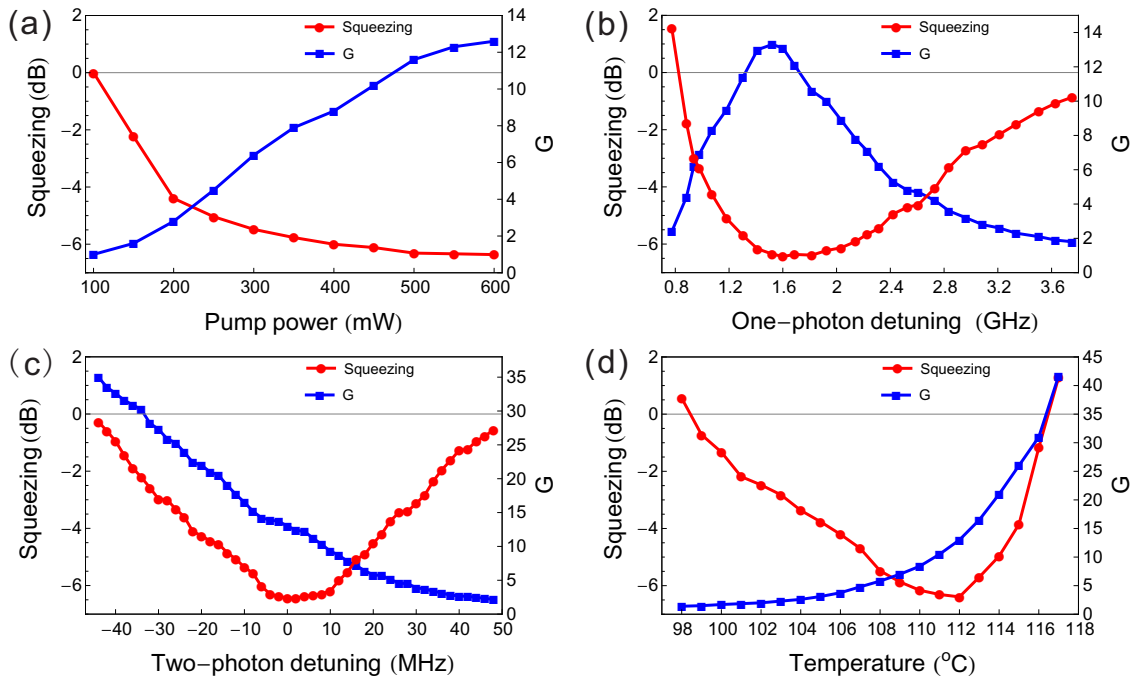}
\caption{Dependence of the normalized intensity-difference squeezing and the FWM gain on (a) pump power, (b) one-photon detuning, (c) two-photon detuning,
and (d) temperature of the vapor cell. The traces for squeezing are plotted in red, and the traces for the FWM gain are in blue. The gray lines at 0 dB show the corresponding SNL.}
\end{figure*}

\section{Results and discussions}
We measure the intensity-difference noise power spectra of the probe and conjugate beams for three different methods to generate the probe beam (indicated as traces $A$, $B$, and $C$ in Fig. 3).
All of these three traces are normalized to the corresponding SNL (trace $D$ in Fig. 3). When two independent lasers are used, the maximum degree of intensity-difference squeezing is 3.7 dB at 0.23 MHz with a squeezing bandwidth of 0.72 MHz (trace $A$).
When the PLL is used, the maximum degree of intensity-difference squeezing is improved to 5.9 dB at 0.23 MHz (trace $B$). However, trace $B$ is even more noisy than trace $A$
in the frequency range from 0.72 to 4 MHz. This may be due to the excess intensity noise induced by the PLL since it modulates the current of the diode laser \cite{LvovskyPLL}.
Trace $C$ shows the intensity-difference noise spectrum when the probe is generated by the EOM sideband. The maximum degree of intensity-difference squeezing is 6.5 dB at around 1 MHz with
a much larger squeezing bandwidth of over 4 MHz. It shows that keeping the frequency and phase difference between the pump and probe beams stable is critical in our experiment and the EOM sideband is an optimal option.

To better compare the noise reduction of the FWM process  using two independent lasers, PLL, and EOM, we vary the seed probe power and record the intensity-difference noise power versus the total optical power of the twin beams (curves $A$, $B$, and $C$, respectively, in Fig. 4).
It must be pointed out that optimal frequency is chosen for each configuration (0.23 MHz for curves $A$ and $B$ and 1 MHz for curve $C$).
Similarly, we also record the noise power of a coherent beam  at 1 MHz at different optical powers using the SNL measurement method described above (curve $D$ in Fig. 4). After fitting these four curves to straight lines, we find that the ratios of slopes between
curves $A$, $B$, and $C$ and curve $D$ are equal to 0.421, 0.262, and 0.222, respectively, which shows that the degrees of intensity-difference squeezing of the twin beams generated in these three configurations are 3.7, 5.8, and 6.5 dB, respectively.

It can be clearly seen that the highest degree and bandwidth of intensity-difference squeezing, among three configurations, can be achieved by using the EOM sideband from Figs. 3 and 4.
We then study the dependence of the FWM gain and the degree of intensity-difference squeezing on the  experimental parameters, such as pump power, one-photon detuning, two-photon detuning, and the temperature of the vapor cell with the probe beam generated by the EOM sideband.

First, the effect of pump power is investigated [shown in Fig. 5(a)]. To do this, we measure the intensity-difference noise power for the twin beams and the corresponding SNL at 1 MHz as a function of pump power and then
calculate the difference of these two mean values of data points on these two curves. For simplicity, we fix $\Delta = 1.6$ GHz, $\delta = 0$ MHz, and $T$ = 112 $^{\circ}$C. It can be seen that both the gain and degree
of intensity-difference squeezing increase with the pump power, in agreement with the expected enhancement of the optical nonlinearity. However, other nonlinear effects (such as self-focusing \cite{SelfFocusing})
set an experimental limit even though the pump power increases further.

Next, we set the pump power at 600 mW, $T$ = 112 $^{\circ}$C, and $\delta = 0$ MHz and scan the one-photon detuning from 0.8 to 3.7 GHz to study the impact of one-photon detuning [shown in Fig. 5(b)].
One can clearly see that quantum correlation represented by the degree of intensity-difference squeezing always exists for most of the region unless \emph{G} is around 1, which shows our system is operated very close to the quantum limit \cite{LettPRA,PooserOE}.
The dependence of the degree of quantum correlation on one-photon detuning can be understood in the following way. Far from resonance, the nonlinearity decreases, which reduces the gain and quantum correlation.
Close to resonance, absorption becomes dominant and thus also degrades quantum correlation.

We then turn our attention to the effect of two-photon detuning when setting pump power at 600 mW, $T$ = 112 $^{\circ}$C, and $\Delta = 1.6$ GHz. As shown in Fig. 5(c), while the FWM gain decreases
all the way as the two-photon detuning varies from $-$44 MHz to $+$48 MHz, the degree of quantum correlation displays an optimum value around 0 MHz. This can be attributed to the resonant enhancement effect.
At higher two-photon detuning, the gain decreases, and thus the degree of quantum correlation decreases. However, at lower two-photon detuning, other higher-order nonlinear effects become dominant,
which also degrades quantum correlation even though the gain increases.

Last, we study the dependence of intensity-difference squeezing on the temperature of the vapor cell [shown in Fig. 5(d)]. We fix the pump power at 600 mW, $\Delta = 1.6$ GHz, and $\delta = 0$ MHz.
Across the presented temperature range (98 $^{\circ}$C to 117 $^{\circ}$C), the gain increases dramatically due to the rapidly changing atomic densities. In contrast, the degree of intensity-difference squeezing displays a maximum value at 112 $^{\circ}$C.
At lower temperatures, the FWM gain is low; thus the quantum correlation between the twin beams is weak. As the temperatures increase further from 112 $^{\circ}$C, the degree of quantum correlation degrades because higher-order nonlinear effects
occur and absorption loss by the hot cesium vapor starts to become more and more dominant.

\section{Conclusion}
In summary, we have measured 6.5-dB intensity-difference squeezing based on the FWM process in hot cesium vapor. Three different methods are used to generate the 9.2-GHz frequency-difference probe beam.
Our result shows that the EOM sideband is an optimal method to achieve a high degree and bandwidth of intensity-difference squeezing. Intensity-difference squeezing can be observed within a wide experimental parameter range, which guarantees its robust generation.
It is expected that a higher degree of intensity-difference squeezing can be achieved by using a shorter vapor cell and higher pump power. This multi-spatial-mode narrowband nonclassical light source near the Cs $D_{1}$ line paves the way to applications involving
quantum imaging and coherent interfaces between light and matter, such as atomic ensembles and a solid-state system.

This research was supported by the Key Project of the Ministry of Science and Technology of China (Grant No. 2016YFA0301402), the National Natural Science Foundation of China (Grants No. 61601270, No. 11474190, and No. 11654002),
the Applied Basic Research Program of Shanxi Province (Grant No. 201601D202006), and the Fund for Shanxi ``1331" Project Key Subjects Construction.

\end{document}